\documentclass{aa}
\usepackage{graphicx}
\begin{document}

%
\title{Discovery of a cataclysmic variable with a sub-stellar companion}
\author{R.E.\ Mennickent \inst{1} \fnmsep\thanks{Based on observations
obtained at the European Southern Observatory, ESO proposals 58.D-0349 and
64.H-0065(B)}
           \and
           M.\ Diaz \inst{2}
           \and
           W.\ Skidmore \inst{3}
           \and
           C.\ Sterken \inst{4}}
   \institute{Departamento de Fisica, Universidad de Concepcion,
Casilla 160-C, Concepcion, Chile. \\ email: rmennick@stars.cfm.udec.cl \and
Instituto Astronomico e Geofisico, Universidade de  Sao Paulo, Brazil. \and
School of Physics and Astronomy, University of St. Andrews, North Haugh,
St. Andrews, Fife, KY16 9SS, UK. \and University of Brussels (VUB),
Pleinlaan 2, 1050 Brussels, Belgium.}

\titlerunning{Discovery of a CV with a sub-stellar companion}
 \authorrunning{Mennickent et al.}

\offprints{R.E.\ Mennickent}
\mail{rmennick@stars.cfm.udec.cl}
\date{}

     \abstract{
We find that the ROSAT source \object{1RXS\,J105010.3-140431} is a
cataclysmic variable with orbital period of 88.6 minutes and a spectrum
closely resembling WZ Sge. In particular, emission lines are flanked by
Stark-broadened absorption wings probably originating in the photosphere
of a compact object. The Balmer absorption lines  can be modeled by the
spectrum of a DA white dwarf with 13\,000 $< T_{eff} <$ 24\,000~$K$.
The strong absorption lines allowed us to obtain direct radial velocities
of the white dwarf using the cross-correlation technique. We find an
extremely low white dwarf radial velocity half amplitude,  $K_{wd}$ = 4
$\pm$ 1 km s$^{-1}$. This is consistent with the upper limit obtained from
the H$\alpha$ emission line wing $K <$ 20 km s$^{-1}$. The corresponding
mass function is incompatible with a  main sequence secondary, but is
compatible with a post  orbital period minimum cataclysmic variable with a
brown dwarf-like secondary. The formal solution gives a secondary mass of
10-20 jovian masses. Doppler maps for the emission lines and the
hypothesis of black-body emission indicate a steady state (T $\sim
r^{-3/4}$) accretion disk mainly emitting in H$\alpha$ and an optically
thicker hotspot with a strong contribution to the higher order Balmer
lines and \ion{He}{I} 5875. As in other long cycle length dwarf novae,
evidence for inner disk removal is found from the analysis of the emission
lines. } \maketitle
%

\section{Introduction}

\object{1RXS\,J105010.3-140431} ($\alpha_{2000}$= +10 50 10.4,
$\delta_{2000}$ = -14 04 32.0) is a poorly-studied ROSAT source detected
with a count rate of 0.06018 $\pm$ 0.01495 counts/sec (Voges et al.\,
1996).
In this paper we present  the first
spectroscopic and photometric study of this object, revealing a probable
classification as a dwarf nova in the
very late stages of the evolutionary track of cataclysmic variables (CVs).

\section{Observations and Data Reduction}

\subsection{Photometric observations}

We observed \object{1RXS\,J105010.3-140431} during three nights on
February 17, 19 and 20, 2000 with the  Danish-1.54m telescope of the ESO
La Silla Observatory. The Danish Faint Object Spectrograph and Camera
DFOSC was used with the backside illuminated Loral/Lesser chip. This CCD
has 2052 $\times$ 2052 15$\mu$m pixels spanning a field of view of
13\farcm3 $\times$ 13\farcm3. A windowing of 5 \arcmin $\times$ 5 \arcmin
was selected around the center to ensure fast readout of the CCD. We
imaged the field of the variable continuously for approximately $\sim$ 2 hours every
night with the  $V$ filter and then measured the instrumental magnitudes of
the variable and comparison stars using the aperture photometry facility
in the software MIRA AP\footnote{MIRA is a registered trademark of Axiom
Research, Inc.}. Differential photometry was performed using nearby
isolated stars in the field. This technique is robust against small
changes in atmospheric transparency and today it is routinely  used for
monitoring variable stars (i.e.\ Howell et al.\ 1988). A finding chart
indicating the comparison stars is shown in Fig.\,1.    The C2 $-$C1 light
curve had a mean value of -1.04 mag and $rms$ 0.017 mag whereas the C3
$-$C2 light curve showed an average of 1.45 mag and $rms$ 0.020 mag. We
did not use the bright star close to the variable as a comparison since it
was saturated in our frames. As C2 has a magnitude similar to the
variable, we deduce an upper limit error of  0.017 mag for the V $-$C2
curve. This light curve, along with the  errors, are shown in Fig.\,2,
suggesting that the star is marginally variable every night at the
2 $\sigma$ level, but the mean magnitude changes from night to night by
0.05 mag.
 
\subsection{Spectroscopic observations}

Spectra were obtained on the 2.2 meter telescope at ESO's La Silla
Observatory, on January 4--7, 1997. The EFOSC2 and CCD \# 24, along with
grism \# 4 and a slit width of 1 arcsecond, yielded a spectral resolution
of 3.4 \AA, as determined by the slit width, plate scale and pixel size,
and a wavelength range of 4020--7470 \AA. He-Ar comparison spectra were
taken  every five  480-s length science exposures. Bias and flat images
were taken daily. Observations are summarized in Table 1.

Data reduction was performed in the usual manner using the IRAF
software package\footnote{IRAF is distributed by the National Optical
Astronomy Observatories, which are operated by the Association of
Universities for Research in Astronomy, Inc., under cooperative
agreement with the National Science Foundation.}.
All of the two dimensional raw frames were de-biased and flat-fielded
using the average bias and normalized flat images, and then one dimensional
spectra were extracted. About 30 He-Ar emission lines provided spectral
calibration functions with typical $rms$ of 0.4 \AA~ (18 km s$^{-1}$ at
H$\alpha$). Flux calibration was performed with the standard stars
\object{LTT\,1020} and \object{EG\,21} (\cite{hamuy92}, \cite{hamuy94}).

\begin{table}
 \caption[]{The spectroscopic observing log indicating starting and ending
times  for the time series and nightly mean   $V$ magnitude.  The number
of science frames per night is also given.}    \begin{center}
\begin{tabular}{|c|c|c|c|c|} \hline \hline \multicolumn{1}{c}{Date(UT)}&
\multicolumn{1}{c}{$N$} & \multicolumn{1}{c}{$HJD_{start}$} &
\multicolumn{1}{c}{$HJD_{end}$}   & \multicolumn{1}{c}{$V$} \\ \hline
05/01/97& 13 &53.7688  &53.8676  &17.63(09)\\ 06/01/97 &14 &54.7928
&54.8188  &17.82(10)\\ 07/01/97 &14 &55.7700  &55.8647  &17.65(05)\\
08/01/97 &17 &56.7590  &56.8704  &17.48(16)\\   \hline \hline
\end{tabular} \end{center} \end{table}

\section{The analysis of the emission line spectrum}

The top of Fig.\,3 shows the averaged spectrum of the object,
characterized by \ion{H}{i} lines and
\ion{He}{i} emission. Flanking the strong
Balmer emission lines we observe the pressure broadened absorption
profiles typical of a white dwarf. We are confident that these lines
come from a white dwarf and not from a hot disk, since
the helium lines are not observed in absorption, like in outbursting dwarf
novae (for a review of spectra during outbursts see Warner 1995,
see also section 4).  The H$\alpha$ emission is double peaked (see bottom
panel of Fig.\,3) with an equivalent width of $-110$\AA~ and half peak
separation 560 km s$^{-1}$. The full width at half maximum $FWHM$ in
H$\alpha$ is 1811 km s$^{-1}$ and the full width at zero intensity $FWZI$
is 2950 km s$^{-1}$. These values are typical for dwarf nova accretion
disks at quiescence. Spectroscopic quantities for the averaged
spectrum are given in Table 2. The steep emission decrement suggests an
origin in an optically thin accretion disk. Although no outburst has been
observed yet for this object, it is likely that it corresponds to a dwarf
nova of the SU UMa subtype. These systems, reviewed by  Warner (1995,
2000),  consist of a white dwarf accreting material from a cool  red dwarf
at very low transfer rates, producing thermal instabilities (outbursts) in
an accretion disk with recurrence times from weeks to several years.  The
faint continuum produced in the disk is hidden in  the brighter white
dwarf spectrum even in  optical wavelengths, producing the absorption
wings seen around the Balmer emission cores. In this regard the spectrum
is quite similar to that of the dwarf nova \object{GW Lib} (Szkody et al.
2000).

\begin{table}
 \caption[]{Mean spectroscopic quantities. Positive equivalent widths
correspond to the absorption component, excluding the emission core.}
\begin{center}
\begin{tabular}{|c|c|c|}  \hline \hline
\multicolumn{1}{c}{Line} &
\multicolumn{1}{c}{flux (erg cm$^{-2}$ s$^{-1}$ \AA$^{-1}$)} &
\multicolumn{1}{c}{W$_{\lambda}$ (\AA)}   \\ \hline
H$\delta$        & 5.9E-16 & -8 \\
H$\gamma$        & 6.0E-16 & -13/+33\\
\ion{He}{I} 4470 & 5.5E-16 & -1 \\
H$\beta$         & 6.5E-16 & -33/+38\\
\ion{He}{I} 5875 & 3.4E-16 & -10 \\
H$\alpha$        & 7.9E-16 & -129  \\
\ion{He}{I} 6678 & 2.0E-16 & -2 \\
  \hline \hline
\end{tabular} \end{center} \end{table}

\subsection{The orbital period and ephemeris}

Derivation of stellar parameters for the binary components in a dwarf
nova requires a precise determination of the radial-velocity curve of at
least one of the stars. This task is not always possible, because in most
systems the white dwarf and secondary star spectrum is hidden in the
optical spectrum of the bright accretion disk. In fact, the emission lines
originating in the disk usually do not reflect the white dwarf binary
motion because they are subject to intrinsic velocity fields besides the
gas rotation in the disk itself  (e.g.\ Robinson 1992). This yields an
intrinsic uncertainty in the spectroscopic ephemeris obtained for most
dwarf novae. The problem has been overcome  in long orbital period dwarf
novae by observing in the infrared the lines of the secondary star.  In
these systems the relative flux contribution and features of the secondary
star allow a straightforward measurement (e.g.\ Friend et al.\ 1990).
However, short orbital period dwarf novae have secondaries too faint to be
detected even in the IR with current instrumentation. In addition, it is
difficult to derive precise radial velocities (R.V.) using molecular
bandprofiles. Fortunately, the disk in \object{1RXS\,J105010.3-140431} is
faint enough to reveal the white dwarf photospheric lines, so we have a
unique opportunity to  obtain white dwarf radial velocities and the
unbiased ephemeris for the system. The spectra were normalized by dividing
them with a low-order polynomial continuum fit before we performed the
radial velocity analysis.

We traced the motion of the H$\beta$ and H$\gamma$
emission line peak by measuring their radial velocity.
Then we constructed periodograms using the  $AOV$
algorithm (Schwarzenberg-Czerny 1989) and the
Scargle (1982) method. We scanned a frequency range
between 90 c/d  (the Nyquist frequency) and 1 c/d. Both methods yielded
consistent results.

The $AOV$ periodogram is shown in Fig.\,4. The main peak is found at
16.272 $\pm$ 0.058 c/d, along with $\pm$ 1 c/d aliases at 15.3 and 17.3
c/d. Based on the strength of the main maximum and assuming a power window
symmetrical regarding the absolute minimum, we  selected 16.272 c/d
(0\fd0615 $\pm$ 0\fd0002) as the most likely period, corresponding
probably
to the binary orbital period. Due to the possible biased
nature of the emission line radial velocities,  we used the H$\gamma$
absorption line (see next section) as the best tracer of the white dwarf
binary motion, measuring the half amplitude of their radial velocity which
resulted basically independent of the period alias, giving confidence about
the stellar parameters derived in Section 3.5.  We found the following
ephemeris for the inferior conjunction of the secondary star:\\

\begin{equation}
 HJD = 2450453.71479(308) + 0.0615(2) E
\end{equation}

\subsection{Comparison between emission and absorption
radial velocities}

In this section we compare the radial velocity sets of emission and
absorption components obtained with different methods.
It was extremely difficult to find reliable velocities for the absorption
component. First we tried extracting the emission core and then fitting a
Gaussian function to the wings, but the line center was poorly constrained
and we obtained noisy results. Secondly we tried deblending the spectral
line with emission plus absorption Gaussians - using the "d" key in the
$splot$ IRAF package, obtaining a
large amplitude for the absorption component (around 110 km$s^{-1}$).
However, numerical simulations showed that this result was an
artifact:  the same apparent oscillation was obtained with a
synthetic zero velocity absorption profile and a moving - 600 km s$^{-1}$
amplitude - emission component (the s-wave, see below). Finally,  we
applied the cross-correlation technique (Tonry \& Davis 1979), implemented
in the $fxc$ IRAF task, to the H$\beta$ and H$\gamma$ white dwarf
absorption,  excluding the emission component inside a range of
-2500 km s$^{-1}$ $< v < $ 2500 km s$^{-1}$. Results for H$\beta$ turned to be too
noisy, but the H$\gamma$ velocity, binned with the orbital period and
fitted with a sine function, allowed us to calculate the time of inferior
conjunction of the secondary star, $T_{0}$ and the radial velocity
half-amplitude: $K_{wd}$ = 4 $\pm$ 1 km$s^{-1}$ (Fig.\,5).

In Fig.\,5 we observe the contrast between  the large
amplitude of the emission line peak and the white dwarf velocity. We will show that the large amplitude
of the emission peak (619 $\pm$ 19 km s$^{-1}$ for  H$\beta$) reflects the
characteristic s-wave observed in the trailed spectra of some dwarf novae
in quiescence. Note that this s-wave moves in antiphase with the white
dwarf, consistent with an origin close to the inner lagrangian point.

Fig.\,5 also shows that the low $K$ amplitude is inconsistent with
noise, as inferred from measurements made on synthetic spectra
with zero velocity  and  noise similar to the
original dataset. We explored the possibility that the low velocity
pattern in Fig.\, 5 is the effect of residual high velocity emission.
However, the s-wave phasing and the disk emission phasing (see below)
indicate that, if present, it should increase the measured amplitude if
coming from the s-wave and should have minor effects if coming
from the disk (the effects are summed or canceled according to the
relative phasing).

The inferred $K_{wd}$ value is amazing. Their accuracy probably
lies in the statistical significance of the cross-correlation technique,
which fits a large sample of pixels simultaneously.  Interestingly, low
$K_{wd}$ values have also been observed in other low mass-transfer rate
dwarf nova which reveal the white dwarf in their optical spectra, namely
\object{WZ Sge} (40 $\pm$ 10 km s$^{-1}$, Spruit \& Rutten 1998),
\object{GW Lib} (40 km s$^{-1}$, Szkody et al.\, 2000) and \object{AL Com}
(0 $\pm$ 32 km s$^{-1}$, Howell et al.\, 1998). All the above values were,
however, obtained from the radial velocity of the emission line wing,
assumed equal to the white dwarf velocity. On the contrary, our results
are the first to directly test  the  white dwarf absorption wings with the
cross-correlation technique.

For H$\alpha$ emission, we applied the ``double Gaussian"
convolution mask algorithm (Schneider \& Young  1980, Shafter 1983, Horne
et al.\ 1986) which provides a robust diagnostic method to investigate the
behavior of different profile sections during the orbital cycle. The method
simultaneously shifts two Gaussians of standard deviation $\sigma_{g}$ (or
alternatively full width at half maximum $FWHM_{g}$) and  center
separation $\Delta$ along the emission profile until a velocity is found
for which the convolved flux in both is the same. Changing $\Delta$ and
$FWHM_{g}$ we can probe different velocity sections of every profile. This
method is more versatile, allowing  us to search for a better trade-off
between noise and velocity resolution by changing $FWHM_{g}$. It is
superior to the simple Gaussian fit for the emission component.

The diagnostic
diagram for the H$\alpha$ emission in Fig.\,6 shows the
variations of $K$ and $\Delta \Phi$ (the phase shift with respect to the
ephemeris defined by the absorption component) for velocity sets obtained
with different Gaussian separations.  We find that $K$ is large close
to the line center but decreases towards the line wing, which could be
the result of a systematic reduction of the s-wave perturbation toward the
line wing.  This is consistent with the diagnostic diagram starting with
Gaussian separations close to the maximum s-wave velocity.
In addition, the $\Delta \Phi$ behaviour indicates that the emission line
center precedes the wing when crossing the zero phase. Interestingly, the
emission line wing is almost in phase with the white dwarf motion.  The
above is consistent with wing emission arising from the inner disk and
following more closely the white dwarf  motion and low velocity emission
coming from a hotspot in the disk-stream impact region close to the inner
Lagrangian point. This view is also consistent with the fact that $K$
 decreases towards the line wing (Fig.\,6 upper panel),
maybe approaching  the white dwarf value, whereas the noise
($\sigma_{K}/K$) becomes prominent.

Phase binned trailed
spectra of the continuum and white dwarf subtracted spectra
(see section 3.5) are shown in Fig.\,7. These spectra confirm the
presence of an additional emission component. We find that this component
is stronger in the higher-order Balmer lines and the \ion{He}{I} 5875
line. This emission component is commonly associated with a hotspot in the
disk and is called the "s-wave". Disk and hotspot are studied in the next
section.

\subsection{Emission line analysis and the optical depth of the disk
and the hotspot}

We measured the Balmer decrement for both the accretion disk and the
hotspot region to check for different line opacities. The Balmer decrement
($D(H\alpha/H\beta$)) is defined as the ratio between the
frequency-integrated specific intensities of the $H\alpha$ and $H\beta$
lines. The fact that the disk and hotspot emission are simultaneously
present in the line is a problem for the calculation of decrements.
De-blending of hotspot and disk emission components is not possible without
a complex physical model for the disk and hotspot regions. We find an
approximate solution for this problem in the following way. We binned the
observed spectra in 10 phase intervals and corrected them for the motion
of the white dwarf. The white dwarf spectrum corresponding to the white
dwarf parameters derived in section 4 was subtracted from these spectra
obtaining phase-binned  emission profiles. To separate the emission
components, we assumed symmetrical disk emission and separated the total
profile for each bin in two components. The first consists mostly of disk
emission, either the red or blue half of the profile depending on the
s-wave velocity, and a second consisting of disk and hotspot emission. The
larger and smaller peak had fluxes $f_{max}$ and $f_{min}$ respectively.
The hotspot decrement was calculated for each phase-binned spectrum
assuming that $f_{hotspot}$ = $f_{max}$ - $f_{min}$ and the disk decrement
assuming $f_{disk}$ = $f_{min}$. We also assumed a similar shape for the
profiles at different wavelengths and applied a correction factor to
calculate the decrements in the $f_{\lambda}$ space.   Evidently the above
is a rough approximation, and phase-bins when the profile was almost
symmetric (i.e.\ the hotspot component was near the middle of the profile)
were not measured.

Results and comparison with theoretical decrements by Williams (1991) are
shown in Fig.\,8. Williams modeled optically thin gas in the emission
lines in accretion disks. He computed H$\beta$ strengths and Balmer
decrements for a grid of temperatures,  inclinations and mid-plane
accretion disk density ($N_{0}$ in units of nucleons per cm$^{3}$).
Fig.\,8 shows the results of his models for a representative  inclination
of 52$^{o}$ along with our observations. Results for different inclination
angles are not significantly different. A direct interpretation of Fig.\,8 is not
possible since the decrements are not single-valued functions of disk
temperature. However,
considering some well-established facts, it is possible to extract some
useful information.
First, the observed disk decrement ($D
(H\alpha/H\beta$) $\approx$  2.3)  indicates that for any realistic disk temperature
($T_{disk}$ $>$ few 10$^{3}$ K) the Balmer lines are optically thin.
Second, the hotspot Balmer decrement  ($D (H\alpha/H\beta$) $\approx$
0.8) is typical of optically thick line emission and, if the lines are
saturated, it would correspond to a black-body temperature of 8000$K$. The
difference between hotspot and disk decrement indicates a disk optically
thinner than the hotspot in the lines. The previous analysis was applied
to the prototype dwarf nova \object{WZ Sge} by Mason et al.\, (2000).
These authors found  values of 3.8 and 2 for the disk and hotspot
decrements respectively, indicating that the disk density in
\object{1RXS\,J105010.3-140431} is not as low as in \object{WZ Sge}.

\subsection{Evidence for inner disk removal}

In the previous section we have shown that the accretion disk of
\object{1RXS\,J105010.3-140431} is similar in some regard to
that found in  long-cycle dwarf novae like \object{WZ Sge}. It has been
suggested that these objects have disks which are depleted in
their inner regions (e.g.\ Hameury et al.\, 2000, Mennickent 1999,
Mennickent \& Arenas 1998). One method to test this scenario for
 \object{1RXS\,J105010.3-140431} is looking into the  emission line
widths. Smak (1981) found that the inner to outer disk radius ratio $R$
and the emissivity exponent $\alpha$ (assuming the emissivity $\propto
r^{-\alpha}$) of an optical and geometrically thin Keplerian disk are
functions of the emission line widths. For  H$\alpha$  we measured 1515,
1915, 2420 km s$^{-1}$ at 80, 40 and 10\% of the maximum intensity level,
respectively, yielding $R \approx$  0.18 and
$\alpha$  $\approx$ 1.5. The emission line widths above were calculated
after subtracting the  underlying absorption profile. The corresponding
figures for the raw emission  are 0.22 and 1.6 respectively. We
observe  \object{1RXS\,J105010.3-140431} close to \object{WZ Sge} in
Fig.\,8 by Mennickent \& Arenas (1998). The rather large $R$ value could
indicate inner disk removal. Using equation (4) of Mennickent \& Arenas
(1998) we calculated a supercycle length of about 5.2 years for
\object{1RXS\,J105010.3-140431}, which should explain the lack of observed
outbursts. This recurrence time should correspond to a mass transfer rate
$\dot{M} \sim$ 4 $\times$ 10$^{14}$ gs$^{-1}$ according to Fig.\,2 by
Warner (1995). This is a very low mass transfer rate  similar only to that
of \object{WZ Sge} ($\dot{M} \sim$ 1 $\times$ 10$^{14}$ gs$^{-1}$) and
\object{HV Vir} ($\dot{M} \sim$ 3 $\times$ 10$^{14}$ gs$^{-1}$, Warner
1995).

\subsection{Synthetic magnitudes and the upper limit for the systemic
inclination}

Synthetic V magnitudes were calculated from the flux-calibrated
spectra. Nightly mean values are given in Table 1. The seeing was
stable during the four observing nights (less than 0.8 arcseconds),
therefore we are confident that the slit losses have little effect
on the mean magnitudes, although they could be the cause of the larger
$rms$ compared with our photometric data. A plot of V versus
orbital phase does not reveal eclipses. This confirms the results of the
differential photometry (Section 2.1).

We can get an upper limit for the inclination in a non-eclipsing CV.
For that we used the Eggleton (1983) approximation
for the volume radius of the Roche lobe of the secondary star,
obtaining:\\

\begin{equation}
\tan{(\frac{\pi}{2}-i_{max})} =
\frac{0.49q^{2/3}}{(1-r_{d}/a)[0.6q^{2/3}+\ln(1+q^{1/3})]}
\end{equation}

\noindent \\
where $q = \frac{M_{2}}{M_{1}}$ is the mass ratio and $a$ the binary
separation. Using the approximation $r_{d}/a$ = $0.6/(1+q)$ (Warner 1995,
Eq.\,2.61), an upper limit $i_{max}$ $\approx$ 65$^{o}$ is found
almost independent of $q$.

\subsection{Constraints on the stellar masses}

The evolution of cataclysmic variables has been recently reviewed
with the use of a population synthesis code by  Howell et al.\,(2001).
The requirement of a Roche lobe-filling secondary allows tracing of the
evolution of the secondary star mass as a function of the orbital period.
For a given orbital period, two solutions  are possible,
depending on whether the system has $\dot{P} < 0$
or $\dot{P} > 0$. From Fig.\, 5 of Howell et al.\, (2001) and using $P=$
88.6 minutes we obtain $M_{2}$ = 0.14  $M_{\sun}$ and $M_{2}$ = 0.05
$M_{\sun}$ respectively. The first solution corresponds to a system with a
nearly main-sequence secondary burning hydrogen in its core. The
system moves to shorter orbital periods as it loses angular momentum by
gravitational radiation. The second solution corresponds to a system with
a secondary star having a hydrogen-exhausted degenerated core, like a
brown dwarf star, and moving to longer orbital periods after "bouncing"
in the orbital period minimum around 80 minutes.
In order to discriminate between both possible scenarios, it is necessary
to use a set of empirical constraints, which we discuss below.

The condition that the
$HWZI$ of the emission lines cannot be greater than the Keplerian velocity
corresponding to the primary radius, combined with the white dwarf
mass-radius relation by Hamada \& Salpeter (1969, yields $M_{1}$ $>$ 0.32
$M_{\sun}$. For this calculation we used the He\,5875 line width ($FWZI$ =
3680 km s$^{-1}$), that probably is not influenced by underlying
absorption.

The low $K_{wd}$
also provides strong constraints on the stellar masses and
systemic inclination through the mass function:\\

\begin{equation}
2 \pi G M_{1} \sin^{3} i = P_{o} K_{wd}^{3} \frac{(1+q)^{2}}{q^{3}}
\end{equation}

\noindent
All the above constraints are plotted in
Fig.\,9. It is evident that a main sequence secondary is incompatible with
the low $K_{wd}$ value.  The formal solution gives a secondary mass
between 10-20 jovian masses for inclination angles between 65 and 35
degree. The inclination would have to be below 5 degrees for a main
sequence companion, which is highly unlikely due to the $FWZI$ of
\ion{He}{I} and the s-wave amplitude. Even if for some
unknown reason the absorption lines do not follow the white dwarf
motion, but the emission wing velocity tends to this value, the mass
function for $K$ = 20 km s$^{-1}$  implies that the inclination should
be lower than 25 degree for a hydrogen burning companion, which is possible
but unlikely.
The mass transfer rate derived in Section 3.4, viz.\,
6.3 $\times$ 10$^{-12} M_{\sun}$ yr$^{-1}$, and the low mass ratio ($q
\leq$ 0.09 derived from the empirical relationship by Mennickent
(1999) using the $K_{wd}/FWHM$ ratio) nicely fit the prediction of Howell
et al.\, (2001) for a post-minimum system with orbital period around 89
minutes.

We have found the first dynamical evidence for
a cataclysmic variable beyond the orbital period minimum. The first
objects belonging to this class, viz.\, \object{LL And} and \object{EF
Eri}, have recently been  discovered by  detecting the
secondary features in the infrared by Howell \& Ciardi (2001),
although indirect evidence has been provided for other objects previously,
namely \object{AL Com} (Howell et al.\, 1998), \object{WZ Sge} (Ciardi et
al.\, 1998), \object{V592 Her} (van Teeseling et al.\, 1999) and
\object{EF Eri} (Beuermann et al.\, 2000).

\subsection{Imaging the accretion disk with Doppler tomography}

The spectroscopic data were phased according to the orbital ephemeris
given above. The continuum was subtracted off in two stages; First, a
cubic polynomial was fit to the continuum regions between the white dwarf
absorption dips and subtracted. Second, a 15000\,K Pop\,II white dwarf
model, convoluted with the instrumental resolution,
and with the continuum removed using a
polynomial as above, was fit to the white dwarf absorption dips in the
data and subtracted to leave only the line emission. Using
maximum entropy Doppler tomography (Marsh \& Horne 1988) we constructed
Doppler maps of the H$\alpha$, H$\beta$, H$\gamma$, H$\delta$ and
\ion{He}{I} 5876 lines. The data were not phase-binned before constructing
the Doppler maps; this maximized phase coverage with 57 spectra covering
all orbital phases observed. The maps are shown in Fig.\,10.  The H$\alpha$
map shows a symmetric disk and a weak hotspot with velocities between
that of the ballistic stream and local Keplerian velocity.
In the H$\beta$, H$\gamma$, H$\delta$
maps the disk becomes progressively fainter, while the hotspot becomes
stronger and shifts to velocities closer to the
local Keplerian velocity. This is consistent with high order
Balmer lines formed deeper in the ballistic stream, where the
excess of kinetic energy is radiated
by interaction with the disk.
The \ion{He}{I} 5876 map shows low velocity bright spot emission; this may
indicate substantial photoionization of the gas stream and/or that the
\ion{He}{I} bright spot originates further up the ballistic stream than
the \ion{H}{I} bright spot, suggesting that the accretion disk extends
further out than would be deduced from the \ion{H}{I} emission alone. In
the nova-like system \object{V347 Pup} for instance, the \ion{He}{II}
emission is produced at inner parts of the disk when compared with
\ion{H}{I} (Diaz \& Hubeny 1999).

The Ratioed Doppler Map for \object{1RXS\,J105010.3-140431} is
shown in the bottom right of Fig.\,10; this was constructed by
dividing the H$\alpha$ map by the  H$\beta$  map shown
in Fig.\, 10. The change in the Balmer decrement
in the region of the bright spot indicates
hotter optically-thick emission than is seen from the disk.
It is interesting to note the similarity in the
structure of the ratioed Doppler maps of \object{1RXS\,J105010.3-140431}
and that of \object{WZ Sge} (Skidmore et al.\, 2000).
However, the ratios of the disk and hot spot
emission measured in \object{1RXS\,J105010.3-140431} are generally lower
than measured in \object{WZ Sge}, indicating higher
temperatures and/or greater optical depth
in \object{1RXS\,J105010.3-140431}, as found in Sec.~3.3.

\subsection{Radial brightness temperature profile}

Using the same method as described in Skidmore et al.\ (2000),
the radial profile of the ratioed Doppler map was calculated by finding
the average ratio at each radius in velocity space. Areas of the ratio map
affected by the bright spot flux were masked when calculating the radial profile.
These areas covered between -20 and 75 degrees measured anti-clockwise from
the y axis. The mean line ratio at each velocity was used to determine a
temperature assuming a simple blackbody source function at the wavelength
of the line emission. This oversimplification is used as a rough guide to
infer temperature gradients in the disk. The system parameters for
\object{1RXS\,J105010.3-140431} determined in this paper were used to
convert the velocity space radial temperature profile into a real space
radial temperature profile. The radial temperature profile for
\object{1RXS\,J105010.3-140431} is shown in Fig.\,11.

We note that the inner disk temperatures are larger than
those suggested by the simple decrement analysis.
Apparently the decrement study gives more weight to the outer disk regions.
This is easily explained since there is a larger contribution of these low
velocity regions to the emission profile and the decrement study is based
on the integrated flux along the profile.

As the slope of the radial temperature in the disk
is close to T $\sim$ $r^{-3/4}$, we  can see that the disk is in a steady
state. In quiescent dwarf novae the disk radial temperature profile is
usually observed to depart from the $r^{-3/4}$ law and to have a much
flatter profile (Wood, Horne \& Vennes 1992, Skidmore et al. 2000). The departure
from T $\sim$ $r^{-3/4}$ occurs at radii of r$_{in}\sim$0.1 and r$_{out}\sim$0.5. If these radii
represent the inner and outer edges of the disk then
R$=\frac{{\rm r}_{in}}{{\rm r}_{out}}\sim$0.2 as found in Sec.~3.4.

\section{Modeling the absorption spectrum}

The  observed  spectrum  of \object{1RXS\,J105010.3-140431}
clearly  shows broad Balmer absorption  profiles
flanking the emission lines.
These features
present  extended  line  wings  typical  of  Stark
broadened Balmer lines; therefore, we may associate
them with the  underlying  white  dwarf  spectrum,
instead of Doppler-broadened  absorption from the
accretion  disk  photosphere.  The conspicuous
presence of the Balmer absorption component indicates that the white
dwarf photosphere is bright when compared to the accretion disk
continuum. In most accreting cataclysmic variable systems the accretion
disk flux prevents the observation of the compact star. The detection
of the white dwarf in    \object{1RXS\,J105010.3-140431} offers
the opportunity of obtaining basic photospheric parameters
for the white dwarf by comparing the observed profiles with a grid
of synthetic spectra. On the other hand, helium absorption features
could not be identified in   the    profiles    of
\ion{He}{I}$\lambda\lambda 6678,5876,4471$ \AA. This fact
suggests a $DA$ classification for the white dwarf.

\subsection{High-gravity   atmosphere  models  and
spectrum synthesis}

In order to obtain a grid of model DA white  dwarf
spectra, pure hydrogen atmospheres were calculated
assuming  hydrostatic  and radiative  equilibrium.
Only  the  continuum  and  hydrogen   opacity  are
included, along with  scattering  and free particle
processes.    The    plane-parallel     atmosphere
structure  is  solved  locally  at  70  mass-depth
points by the complete  linearization method using
the code TLUSTY (Hubeny \& Lanz 1995).

Once  the  density,  temperature  and  opacity  is
calculated,  the  radiative  transfer  equation is
solved  for  each   frequency  in  the  region  of
interest by the program SYNSPEC  (Hubeny, Lanz \&
Jefferys  1994).  NLTE  departures are allowed for
the hydrogen levels population. The improved
calculations   of   the   Stark    broadening   by
Shoning  \&  Butler  (1989) in  addition  to the
thermal Doppler  component are used to compute the
Balmer   line   profiles.   Both   emergent   flux
($H_{\lambda}$)  and  theoretical   continuum  are
computed  in  absolute  units.  Our  line  profile
calculations  were  found to be in good  agreement
with the white  dwarf  models by Jordan \& Koester
(1986).

\subsection{Model    fitting   and   white   dwarf
parameters}

About  90  model   spectra  were   computed   with
effective  temperatures  ranging  from  10000 K to
65000 K and $\log{g}$  between 6.0 and 9.5.  Continuum
subtracted  and  continuum  normalized model  sets  were
prepared using the theoretical
continuum calculated in the absence of bound-bound transitions.
Subsequently we fitted a low order polynomial
pseudo-continuum  to the models in the  region  covered  by our
spectroscopy.  This residual continuum  correction
of model spectra is necessary for their comparison
with the  observations because of the  significant
overlapping   of  the  Balmer  line  wings  beyond
H$\beta$.

The regions for fitting were carefully selected in
the observed  spectrum to exclude the contribution
from the  narrow  central  emission  and from weak
emission   lines  in  the   extended   wings.  Two
different  procedures  were  employed to fit model
spectra to the observations:

The first method
consists of fitting the line profile in flux units
using  continuum-subtracted  spectra.  Each model
spectrum in the grid is arbitrarily scaled to match
the  integrated  flux in the  observed  absorption
lines and then the reduced chi-square statistic is
computed for each pair ($T_{eff}$,  $\log{g}$ ).  This
method does not  require a previous  knowledge  of
accretion disk continuum contribution to the total
flux.  However,   it  has  the   disadvantage   of
disregarding  the line to  continuum  flux  ratio.
Results  from this method are shown in Fig.\,12.  A
valley in the $\chi^{2}$  surface is evident from the
graph,  containing  two  minima.  A sample
individual   fit   can   be   seen   in   Fig.\,13.

A caveat related to the effect of the unknown
interstellar extinction should be made here.  This
effect,  in  principle, must   be   considered   when  the
simultaneous  fitting of  H$\alpha$,  H$\beta$ and
H$\gamma$  is  performed.  However,  a  comparison
with the fit  of H$\beta$ and  H$\gamma$  alone
does  not  show   significant   discrepancies.  In
addition,  the expected  extinction  at the target
galactic latitude  ($b_{II}$ = $39\deg$) should be
low.

The   second   method   uses  the   continuum-normalized  spectrum  as
input and deals  with the veiling  by the  other  continuum  sources  in
the binary.  Unfortunately, the shape of the  underlying disk  continuum
emission  is  unknown.  However, given its low  luminosity  relative  to
the  white dwarf one may expect under standard  theory that such  a  disk
should   present   relatively   low temperatures.  For the sake of
simplicity  it was assumed that the disk  continuum  shape is similar to
the white  dwarf  continuum,  but  scaled  by an unknown factor.  The
relative disk contribution to the  total  flux  ($f$)  is  independent
of  the wavelength and may be left as a free  parameter in the model
fitting.  Typical  results are shown in Fig.\,14, where the behavior is
similar to that seen in Fig.\,12. Simulations show that the minima drift
toward  lower  temperatures and higher gravities as the veiling factor $f$
increases.  The increase in $f$ also implies a significant decrease  in the
 quality  of the fit.  It  became clear from our  simulations  that the
disk  should
contribute less than 20\% of the system flux.

The dynamical solution discussed in Section 3.6
gives a range for the white dwarf mass too large to discriminate between
the two minima found in the $\chi^{2}$ plots.
By  combining the quality criteria  obtained in both procedures we estimate
 $\log{g}$  ranging from 7.6 to 9.3 and $T_{eff}$  from 13\,000 K to
24\,000 K.

\section{Conclusions}

\begin{itemize}

\item A spectroscopic analysis of the ROSAT target
\object{1RXS\,J105010.3-140431} has been realized noting the close
similarity between this source and the prototype dwarf nova \object{WZ
Sge}. In particular, the white dwarf photospheric lines are revealed in
the spectrum, which is a signature of a very low mass accretion rate;

\item We find a most likely orbital period of 88.6 minutes although
$\pm$ 1 cycle/day aliases cannot be ruled out;

\item The absorption spectrum can be modeled
with a DA white dwarf with $13\,000~K < T_{eff} < 24\,000~K$;

\item The mass function suggests that the system is beyond the orbital
period minimum having an undermassive secondary;

\item As in other short orbital period dwarf novae systems, evidence
for inner disk removal is
found. In addition,  using empirical
relationships for dwarf novae,  we predict a supercycle of several years
for this object;

\item The Doppler maps and the Balmer decrement study reveal significant
differences between the optical depth of the disk and hotspot.
We find an optically thick hotspot with $T \sim$ 8000 $K$ and an
optically thinner disk. The hotspot has velocities consistent with
dissipation of kinetic energy in the stream-disk interacting region,
and it is the main emitting source in high order Balmer lines and the
\ion{He}{I} 5875,  but not in H$\alpha$, where the emission is mainly
distributed along a symmetrical disk. Contrary to the Balmer lines, the
helium line seems to be formed in the outer stream-disk impact region;

\item We constructed the temperature profile for the disk and, contrary to
that found in other dwarf  novae, we find an almost steady state
standard disk, with T $\sim$ $r^{-3/4}$.

\end{itemize}

\begin{acknowledgements}

This work was partly supported by
Fondecyt 1000324, D.I. UdeC 99.11.28-1 and
Fundacion Andes C-13600/5. This project was also supported by the
Flemish Ministry for Foreign Policy, European Affairs, Science and
Technology and by FAPESP 9906261 and CNPq 301029. We thank Thomas
Augusteijn for calling our attention to this interesting object during an
observing  session in the La Silla Observatory. We also thank Tom Marsh
for the use of his {\bf MOLLY} spectral analysis software and {\bf
DOPPLER} maximum entropy Doppler tomography software. We thank Detlev
Koester for providing the white dwarf spectrum used during the production
of the Doppler maps. We also thank Ligia Barros for her kind help during
the photometric reductions.

 \end{acknowledgements}

{\bf Figure Captions}
\begin{itemize}

\item Fig.1 An ESO Digitised Sky Survey image indicating the comparison
stars used for differential photometry. The field of view is 5\arcmin
$\times$ 5\arcmin.

\item Fig.2 The differential magnitudes in the $V$ band along with their
error bars.

\item Fig.3  The averaged spectrum (upper panel) and a close up view
of the main spectral lines (below panel). From up to down the
\ion{He}{5875}, H$\gamma$, H$\beta$ and H$\alpha$ lines are shown.

\item Fig.4 The $AOV$ periodogram of the H$\beta$ emission line. The
maximum at 16.27 c/d (0\fd0615) is identified as the orbital period.

\item Fig.5 Upper panel: The H$\beta$ (squares) and H$\gamma$
(circles) s-wave velocity folded with the ephemeris given by Eq. 1. Below
panel: The H$\gamma$ absorption line velocity (squares) obtained
by cross correlation and the best sine fit along with results for zero
velocity synthetic absorption profiles (circles).

\item Fig.6 The diagnostic diagram for the H$\alpha$ emission line
considering the ephemeris given by Eq.\,(1).

\item Fig.7 The trailed spectra of the emission lines. The data have been
binned into 20 phase bins and are repeated for clarity. Note how the
relative strength of the bright spot emission with respect to the disk
emission increases from H$\alpha$ to H$\delta$. A double peak is evident
in H$\alpha$.

\item Fig.8 Theoretical Balmer decrements for optically thin disks emission
lines for inclination of 52$^{o}$ according to Williams (1991). The models
for T= 8000 K, 10\,000 K, and 15\,000 K, all approach to
D(H$\alpha$/H$\beta$) = 1 at high densities, as expected for optically
thick lines. The lines indicate orbit-averaged decrements
for disk and hotspot and their corresponding uncertainties.

\item Fig.9 The $M_{1}-M_{2}$ plane showing the dynamic solutions for
$K_{1}$ = 4 km s$^{-1}$ (absorption wings) and $K_{1}$ = 20 km s$^{-1}$
(emission wings).  The vertical lines indicate possible secondary star
masses for the orbital period accordingly to results of the evolutionary
code by Howell et al.\, (2001). A post orbital period minimum secondary
is favoured by our velocities.

\item Fig.10 Doppler maps of the H$\alpha$, H$\beta$, H$\gamma$, H$\delta$
and \ion{He}{I} 5876 emission lines. Overlayed is the Roche geometry of
the system with $M_{1}$ = 1.0 $M_{sun}, M_{2} = 0.05 M_{sun}$,
P$_{orb}$ = 0.0615 d and i = 65 deg. The dashed line shows the local
Keplerian velocity along the stream path assuming circular orbits around
the primary. The circular line represents 1/3 of the primary Roche
lobe radius. The relative increase in bright spot strength with respect to
the disk emission is evident as we go from H$\alpha$ to H$\delta$. Note
that the \ion{He}{I} bright spot is positioned with a slower velocity
than the \ion{H}{I} bright spots. In general the bright spots have velocity
between that of ballistic stream and the local Kepler velocity.  Bottom
Right: The Ratio map of the H$\alpha$ and H$\beta$ Doppler maps. Note the
striking similarity in the structure of the ratio map of
\object{1RXS\,J105010.3-140431}  to that of \object{WZ Sge} shown in
Fig.\, 6 of Skidmore et al.\, (2000).

\item Fig.11 The radial temperature profile for the disk emission in
\object{1RXS\,J105010.3-140431} assuming a blackbody source function. The
diagonal dashed line indicates the slope of the radial temperature profile
for a standard alpha disk where T $\sim$ $r^{-3/4}$.

\item Fig.12 $\chi^{2}$ as a function of surface gravity and temperature
according to the results of Method 1.

\item Fig.13 Illustration of the spectrum synthesis analysis.

\item Fig.14 $\chi^{2}$ as a function of surface gravity and temperature
according to the results of method 2. The plot corresponds to a value of
$f$ = 0.08 (see text).

\end{itemize}

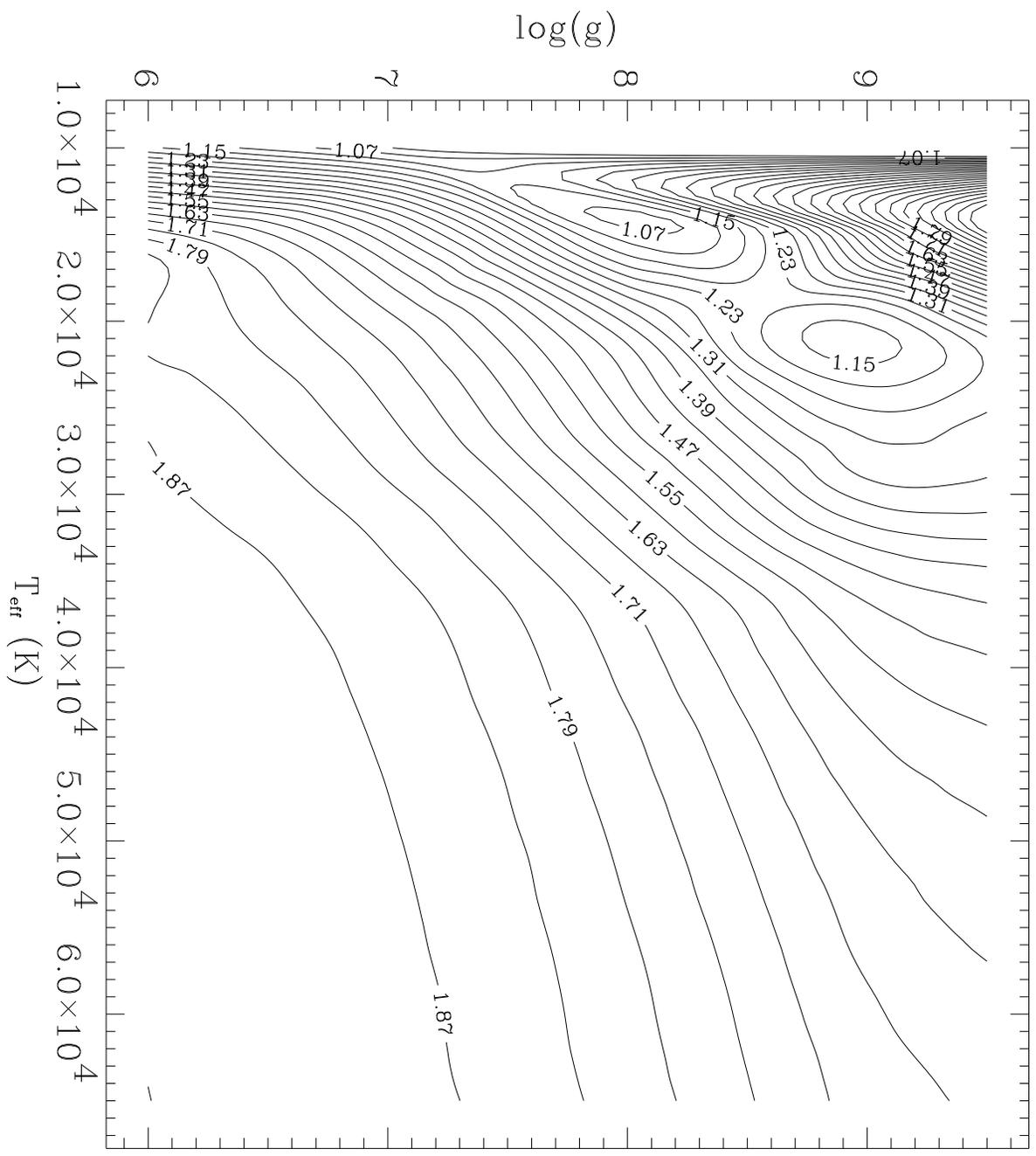
\end{document}